\documentclass{elsart}
\usepackage{epsfig}

\newcommand{\ba}{\begin{eqnarray}}
\newcommand{\ea}{\end{eqnarray}}
\newcommand{\bas}{\begin{eqnarray*}}
\newcommand{\eas}{\end{eqnarray*}}
\newcommand{\be}{\begin{equation}}
\newcommand{\ee}{\end{equation}}
\newcommand{\bes}{\begin{equation*}}
\newcommand{\ees}{\end{equation*}}

\newcommand{\bcentre}{\begin{center}}
\newcommand{\ecentre}{\end{center}}

\font\tenmsb=msbm10 scaled\magstep1
\font\sevenmsb=msbm7 scaled\magstep1
\font\fivemsb=msbm5 scaled\magstep1
\newfam\msbfam
\textfont\msbfam=\tenmsb
\scriptfont\msbfam=\sevenmsb
\scriptscriptfont\msbfam=\fivemsb

\newcommand{\order}[1]{{\mathcal O}(#1)}

\begin{document}    
\begin{flushright}\begin{small}
Edinburgh/2002/01\\
\end{small}\end{flushright}
\vskip1ex
\begin{frontmatter}
\title{Semi-leptonic decays of heavy mesons and the Isgur-Wise function in quenched lattice QCD}
\collab{UKQCD Collaboration}
\author[ed]{K.C.~Bowler}, 
\author[ed]{G.~Douglas\thanksref{douglas}},
\author[ed]{R.D.~Kenway},
\author[ed]{G.N.~Lacagnina},
\author[ed]{C.M.~Maynard}
\thanks[douglas]{CCIR, University of Edinburgh, 80 South Bridge, Edinburgh EH1 1HN, Scotland, UK}
\date{\today}
\address[ed]{Department of Physics \& Astronomy, University of Edinburgh,
        Edinburgh EH9 3JZ, Scotland, UK}
\date{\today}

\begin{abstract}
The form factors for the semi-leptonic $B\to D$ and $B\to D^*$ decays
are evaluated in quenched lattice QCD at two different values of the
coupling, $\beta=6.0$ and $6.2$. The action and the operators are fully
$\mathcal{O}(a)$ non-perturbatively improved. The slope of the
Isgur-Wise function is evaluated, and found to be
$\rho^2=0.83^{+15+24}_{-11-1}$ (quoted errors are statistical and
systematic respectively). Ratios of form factors are evaluated and
compared to experimental determinations.
\end{abstract}
\end{frontmatter}


\section{Introduction}

The $B\to D$ and $B\to D^*$ semi-leptonic decays are of considerable
phenomenological interest. They provide ample opportunity for
interaction between experiment and theory. The ideas of Heavy Quark
Symmetry
(HQS)~\cite{HQS_russian1,HQS_russian2,HQS_russian3,wisgur1,wisgur2}
and Heavy Quark Effective Theory (HQET)~\cite{georgi} were first
developed and applied to these decays. A combination of theoretical
and experimental data for semi-leptonic decays can be used to
determine the CKM matrix element $|V_{cb}|$. HQS allows
theoretical control of the non-perturbative (NP) aspects of the
calculation around the infinite quark mass limit. In this study we
determine the non-perturbative matrix elements directly on the
lattice.

HQS can be used to constrain the form of the matrix elements that
describe semi-leptonic decays of heavy-light mesons. In particular,
the relevant matrix elements are expressed in terms of a set of form
factors that contain the non-perturbative physics of the decay:

\begin{eqnarray}
\frac{\langle D(v')|{\bar c}\gamma^\mu b|B(v)\rangle}{\sqrt{m_B m_D}} & = & (v+v')^\mu h_{+}(\omega)+(v-v')^\mu h_{-}(\omega)\\
\frac{\langle D^*(v',\epsilon)|{\bar c}\gamma^\mu b|B(v)\rangle}{\sqrt{m_B m_{D^*}}} & = & i\epsilon^{\mu\nu\rho\sigma}\epsilon_{\nu}^{*}v_{\rho}'v_{\sigma}h_V(\omega)\\ 
\frac{\langle D^*(v',\epsilon)|{\bar c}\gamma^\mu\gamma^{5} b|B(v)\rangle}{\sqrt{m_B m_{D^*}}} & = & (\omega+1)\epsilon^{*\mu}h_{A_1}(\omega)+\nonumber \\& & -[h_{A_2}(\omega)v^\mu+h_{A_3}(\omega)v'^{\mu}](\epsilon^*\cdot v)
\end{eqnarray} 

where $v,\ v'$ are the velocities of the initial and the final meson
respectively, and $\omega=v\cdot v'$; $\epsilon$ is the polarisation
vector of the $D^{*}$ meson. The $\omega$ variable is kinematically
constrained to the interval

\begin{equation}
1 \le \omega \le \frac{m_B^2+m_{D^{(*)}}^2}{2m_Bm_{D^{(*)}}}\ .
\end{equation}

In the limit of infinitely heavy bottom and charm quarks, the six form
factors are related to a universal function known as the Isgur-Wise
function $\xi(\omega)$~\cite{wisgur1,wisgur2}. Away from the heavy
quark limit this relation is modified by two kinds of corrections:
perturbative QCD corrections and heavy quark symmetry breaking
corrections. For large enough heavy quark masses, the relationships
between the form factors and $\xi(\omega)$ are

\begin{equation}
h_{j}(\omega)=[\alpha_j+\beta_j(m_b,m_c;\omega)+\gamma_j(m_b,m_c;\omega)+
	\order{1/m_{b,c}^2}]\xi(\omega)\ .
\end{equation}

The $\alpha_j$ terms are constants that fix the behaviour of the form
factors in the heavy quark limit
($\alpha_{+}=\alpha_V=\alpha_{A_1}=\alpha_{A_3}=1$;
$\alpha_-=\alpha_{A_2}=0$). The $\beta_j$ and $\gamma_j$ functions
account respectively for radiative corrections and power corrections
proportional to the inverse of the heavy quark mass. The radiative
corrections are calculable in perturbation
theory~\cite{neubert_wisgur_rad}, while the power corrections are
non-perturbative in nature
\cite{neubert_power}. At zero recoil ($v=v',\omega=1$) however,
Luke's theorem \cite{luke_sub} guarantees that $\gamma_+=0$
and $\gamma_{A_1}=0$, which means that power corrections to
$h_+$ and $h_{A_1}$ are of order ${\mathcal O}(1/m_{b,c}^2)$. 

At small recoil, the Isgur-Wise function is modelled by

\begin{equation}
\label{eqn:linearwisgur}
\xi(\omega)=1-\rho^2(\omega-1)+{\mathcal O}((\omega-1)^2)\ 
\end{equation}

where $\rho^2$ is called the slope parameter and $\xi(1)=1$, because of
current conservation. Alternative parametrisations of $\xi(\omega)$
are possible, which start to differ from (\ref{eqn:linearwisgur}) at
${\mathcal O}((\omega-1)^2)$ \cite{wisgur_model_1,wisgur_model_2}.

In this paper we present a study of the form factors of the $B\to D$
and $B\to D^*$ semi-leptonic decays, and the extraction of the
Isgur-Wise function in the quenched approximation to lattice QCD. The
calculations are performed at two values of the coupling, $\beta=6.0$
and $\beta=6.2$, using a non-perturbatively improved relativistic
Sheikholeslami-Wohlert (SW)~\cite{sw_paper} fermion action and current
operators, so that the leading discretisation errors appear at
${\mathcal O}(a^2)$ rather than ${\mathcal O}(a)$, where $a$ is the
lattice spacing. Whilst the improvement gives better control over the
continuum extrapolation, it does not necessarily reduce the size of
lattice artifacts at fixed coupling.  With only two values of the
coupling, no continuum extrapolation is attempted.

The remainder of the paper is organised as follows. In section $2$ we
give details of the calculation including extraction of the form
factors. In section $3$ we describe our determination of the vector
and axial current renormalisation constants. In section $4$ we report
the results of the extraction of the Isgur-Wise function and its
slope. Section $5$ dwells on the problem of quark mass dependence,
while section $6$ describes a calculation of ratios of form factors
and its comparison with experimental determinations. An analysis of
the systematic sources of uncertainty and a comparison with other
results is detailed in section $7$.

\section{Details of the calculation}

\subsection{Improvement of action, currents and masses}

The improved action used in this work has the form

\begin{equation}
S_{\mathrm{SW}}=S_{\mathrm{W}} - c_{\mathrm{SW}} \frac{i\kappa}{2}\sum_{x}
{\bar \psi}(x) i \sigma_{\mu\nu} F_{\mu\nu}(x) \psi(x) 
\end{equation}

where $S_{\mathrm{W}}$ is the Wilson action. The improvement programme
also requires the improvement of current operators; in particular, the
improved vector and axial currents are

\begin{eqnarray}
\label{eqn:imp_current}
V_{\mu}^\mathrm{I}(x) & = & V_{\mu}(x)+ac_V{\tilde
                         \partial}_{\nu}T_{\mu\nu}(x) \nonumber \\
A_{\mu}^\mathrm{I}(x) & = & A_{\mu}(x) +
                         ac_A{\tilde \partial}_{\mu}P(x)
\end{eqnarray}
where
\begin{eqnarray}
V_{\mu}(x) & = & {\bar \psi}(x) \gamma_{\mu} \psi(x) \nonumber \\
A_{\mu}(x) & = & {\bar \psi}(x) \gamma_{\mu}\gamma_{5} \psi(x)\nonumber \\
P(x)       & = & {\bar \psi}(x) \gamma_{5} \psi(x) \nonumber \\
T_{\mu\nu}(x) & = & {\bar \psi}(x)i\sigma_{\mu\nu} \psi(x) \nonumber
\end{eqnarray}

and ${\tilde \partial_{\mu}}$ is the symmetric lattice derivative in
the ${\hat \mu}$ direction, defined in terms of the lattice spacing $a$ by

\begin{equation}
\tilde \partial_{\mu}f(x)=\frac{1}{2a}[f(x+a{\hat \mu})-f(x-a{\hat \mu})]\ .
\end{equation}

The generic renormalisation of improved current operators is performed
as follows $(J=A,V)$:

\begin{equation}
\label{eqn:renorm_current}
J^\mathrm{R} = Z_J(g^2)(1+b_J(g^2)am_q)J^\mathrm{I}
\end{equation}

where $Z_J$ is calculated in a mass-independent renormalisation
scheme. It is conventional to define effective renormalisation
constants:

\begin{equation}
\label{eqn:eff_renorm}
Z_J^{\rm eff} = Z_J(1+b_Jam_q)\ .
\end{equation}

The bare quark mass, in terms of hopping parameters, is equal to

\begin{equation}
\label{eqn:amq}
am_q=\frac{1}{2}\left(\frac{1}{\kappa_q}-\frac{1}{\kappa_{\rm crit}}\right)
\end{equation}

where $\kappa_{\rm crit}$ is the value of the hopping parameter at
which the bare quark mass vanishes. The improved, renormalisation
group invariant (RGI) quark mass is defined in the following way

\begin{equation}
\label{eqn:quark_mass}
a{\tilde m}_q= Z_m am_q(1+b_m am_q)\ .
\end{equation}

A discussion of the different determinations of the improvement
coefficients can be found in~\cite{np_imp_fb}. We use the NP value of
$c_{\rm SW}$ determined by the ALPHA
collaboration~\cite{alpha_np,alpha_np2} and the improvement
coefficients for the current operators are taken from Bhattacharya
{\em et al.}~\cite{Bhatta_99,bhatta_plb,bhatta_2000}, except for $Z_m$ and
$b_m$. The additive $b_m$ coefficient is evaluated at 1-loop in perturbation
theory~\cite{alpha_np4_bA} with the coupling ``boosted'' by the mean link,
$g^2\to g^2/u_0^4$, and $Z_m$ is determined non-perturbatively by the ALPHA
collaboration~\cite{alpha_ZM}.

\subsection{Simulation details}

The lattice spacing is set using the Sommer scale $r_0$
\cite{sommer,wittig_r0}.  The matrix elements are extracted from
combinations of quark propagators corresponding to four values of the
heavy quark masses ($m_{\rm heavy}\simeq m_{\rm c}$) and two values of the mass
of the light (passive) quark ($m_{\rm light}\simeq m_{\rm s}$). Simulation details
are summarised in Table
\ref{tab:sim_details}. The gauge configurations that are used in
this calculation were generated with a combination of the over-relaxed
\cite{creutz_or,brown_woch_or} and the Cabibbo-Marinari
~\cite{cabibbo_marinari} algorithms with periodic boundary
conditions. The light quark propagators are smeared with the fuzzing
technique \cite{fuzzing}, while heavy quark propagators are smeared
using a gauge invariant technique \cite{boyling_p}. The statistical
errors are estimated using the bootstrap method
\cite{efron}.

\begin{table}
\begin{center}\caption{Simulation details.\smallskip}
\label{tab:sim_details}
\begin{tabular}{@{}llcc}
\hline
&& \multicolumn{1}{c}{$\beta=6.2$}
 &\multicolumn{1}{c}{$\beta=6.0$}\\
\hline
Volume && $24^4\times 48$ & $16^3\times 48$ \\
$N_{\rm configs}$ && $216$ & $305$ \\
$c_{\mathrm SW}$ && $1.614$ & $1.769$\\
$a^{-1}(r_0)({\rm MeV})$ && $2.913$ & $2.123$\\
$\kappa_{\rm heavy}$ && $0.120, 0.1233, 0.1266, 0.1299$ & $0.1123,0.1173,0.1223,0.1273$\\
$\kappa_{\rm light}$ && $0.1346, 0.1351$ & $0.13344,0.13417$\\
\\
\hline
\end{tabular}
\end{center}
\end{table}

\subsection{Extraction of the form factors}

The matrix elements relevant to the semi-leptonic decays are extracted
from fits to two and three-point correlation functions. The general
form of the latter is as follows

\begin{equation}
C^\mu_{3{\rm pt},J}(\vec{p}_A,t_x;\vec{p}_B,t_y)=\sum_{\vec{x},\vec{y}} e^{-i(\vec{p}_B\cdot\vec{x}+\vec{q}\cdot\vec{y})}\langle 0 | T[\Omega_B(x)J^{\mu\dagger}(y)\Omega_A^{\dagger}(0)]|0\rangle 
\end{equation}

where $\vec{q}=\vec{p}_B-\vec{p}_A$ and $\Omega_X^\dagger$ is the
operator that creates the state $X$. $J^\mu$ is the weak current. We
have computed the correlation function using the ``standard source''
method~\cite{ext_prop,ext_prop2}.  The time-slice $t_x$ is fixed to
$t_x=28$, i.e. slightly off the centre of the lattice, in order to
allow for an analysis of the different systematic effects of the two
sides of the lattice. The {\em extended} heavy quark propagators 
were computed at four values of the hopping parameter. For the active
quark propagators, which are part of the current, only two values
of the hopping parameter were used at $\beta=6.2$ due to disk space
constraints and all four were used at $\beta=6.0$.

\begin{figure}[!ht]
\begin{center}
\epsfig{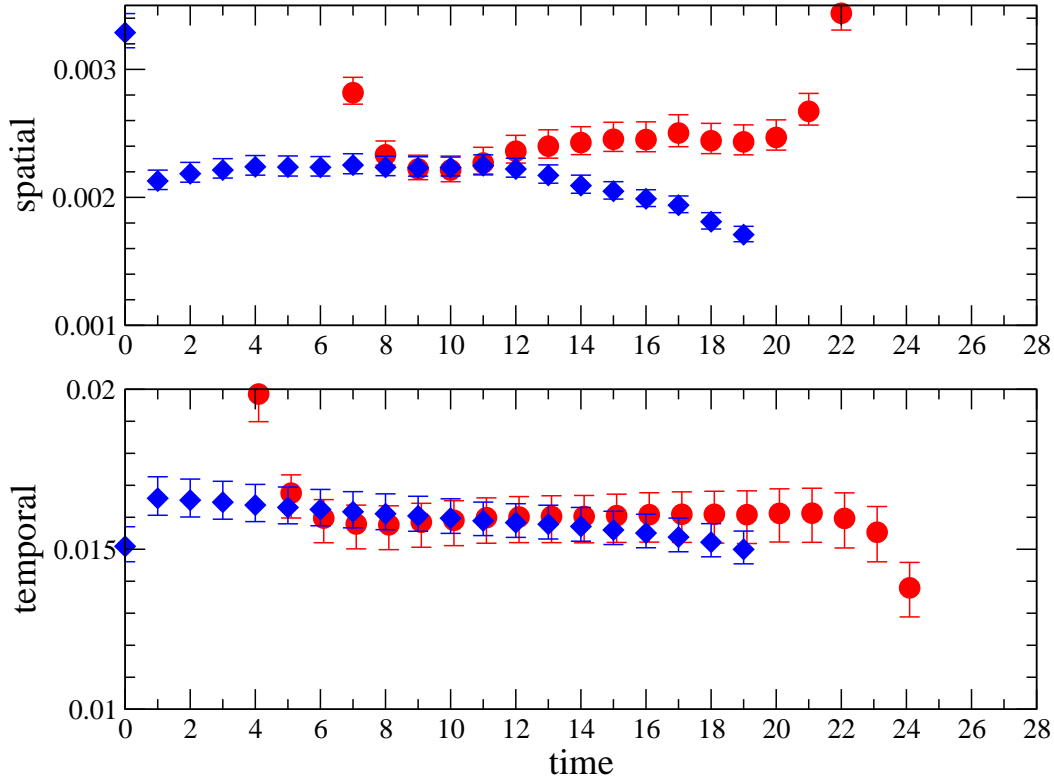}
\caption{The matrix element 
	$\langle P_B(\vec{p}_B) | V_\mu | P_A(\vec{p}_A)\rangle$ 
	with $\beta=6.2$, $\kappa_H=\kappa_{H^\prime}=0.1200$,
	$\kappa_L=0.1346$, $|\vec{p}_A|=0$ and $|\vec{p}_B|=1$.
	The data shows the ratio of the three-point function over the
	fitted two-point parameters. The circles are $0<t_y<27$ (fore side)
	and the diamonds are $0<(T-t_y)<19 $ (back side).}
\label{fig:matrix_element}
\end{center}
\end{figure}

In the Euclidean time formulation, when the
time separation of the operators is large enough, i.e. when $t_y$ and
$t_x-t_y$ are large, one has

\begin{equation}
C^\mu_{3{\rm pt},J}(\vec{p}_A,t_x;\vec{p}_B,t_y)=\frac{Z_A}{2E_A}e^{-E_At_y}\frac{Z_B}{2E_B}e^{-E_B(t_x-t_y)}\langle B(\vec{p}_B)|J^{\mu \dagger}(0)|A(\vec{p}_A)\rangle
\end{equation}

where $Z_i=\langle 0|P_i|P_i(\vec{p}_i)\rangle$. The $Z_i$ constants
and the energies $E_i=\sqrt{|\vec{p}_i|^2+m_i^2}$ are extracted from
fits of two-point functions (detailed descriptions of the methods
employed can be found in
\cite{np_imp_fb,QLHS}). The $\{Z_i,E_i\}$ parameters are then used to cancel the asymptotic
time dependence from the three-point function and to extract the
desired matrix element from a fit to a plateau. For each value of
$\omega$ where
\begin{equation}
\omega=\frac{m_B^2+m_{D^{(*)}}^2-q^2}{2m_Bm_{D^{(*)}}}
\end{equation}
an average is performed of all the kinematic and Lorentz channels for
which the matrix elements have the same value. An example matrix
element is shown in figure~\ref{fig:matrix_element}. All distinct
matrix elements are fitted simultaneously to extract the form factors.
We examine five kinematic channels, each of them being specified by
the values of $|\vec{p}_A|,|\vec{q}|$. These are summarised in Table
\ref{tab:kin_chans}.

Since the heavy quarks used in this study have masses around that of
the charm quark, we are in effect extracting the Isgur-Wise function
from a $D\to D^{(\star)}$ decay. Provided that there is little
residual mass dependence beyond leading order power and radiative
corrections in the form factors, then the machinery of HQS allows us to
compare to experimental $B\to D^{(\star)}$ decays.

\begin{table}[!ht]
\begin{center}
\caption{Spatial moduli of momenta in the different kinematic channels, in 
	lattice units of $2\pi/(aL)$. \smallskip}
\label{tab:kin_chans}
\begin{tabular}{ccc}
\hline
 Channel\ \# & $|\vec{p}_A|$ & $|\vec{q}|$ \\\hline $0$ &$0$&$0$\\ $1$
 &$0$&$1$\\ $2$ &$1$&$1$\\ $3$ &$1$&$0$\\ $4$ &$1$&$\sqrt{2}$ \\\hline
\end{tabular}
\end{center}
\end{table}

\section{Current renormalisation}

\subsection{The vector current}
Observing that the degenerate mass vector current is conserved in the
continuum, we consider the effective matching coefficient $Z_V^{\rm
eff}$, obtained from the forward matrix element of the current between
degenerate heavy-light mesons at rest. We then have
\begin{equation}
\label{eqn:ZV_eff}
  \langle P(\vec{0})\; | V_0 | P(\vec{0})\; \rangle^{\rm cont}
	= 2 M  = Z_V^{\rm eff} 
     \langle P(\vec{0}) \;| V^I_0 | P(\vec{0}) \;\rangle^{\rm latt}\ .
\end{equation}
In particular, taking the ratio of a heavy-light two-point function,
at the extension time-slice, with the forward degenerate matrix element 
gives,
\ba
  \frac{C_{2\rm pt}(\vec{p}=\vec{0};t=t_x)}
	{C^{A=B}_{3\rm pt}(\vec{p_A}=\vec{p_B}=\vec{0};t_x,t_y)} & = &
	\frac{Z^2e^{-Mt_x}}{2M}\times \\\nonumber
	& &\frac{(2M)^2}
	{\langle P(\vec{0}) \;| V^I_0 | P(\vec{0}) \;\rangle^{\rm latt}
	Ze^{-Mt_y}Ze^{-M(t_x-t_y)}} \\\nonumber
	&=&\frac{2M}
	{\langle P(\vec{0}) \;| V^I_0 | P(\vec{0}) \;\rangle^{\rm latt}}
	\\\nonumber
	&=& Z_V^{\rm eff}\ .
\ea

The results as a function of bare quark mass, as it appears in
equation (\ref{eqn:eff_renorm}), are given in Table \ref{tab:ZV_VS_M}.
Also shown in the table is the spectator quark mass. We find that the
current renormalisation does not depend significantly on the spectator
quark mass. The results for $\beta=6.2$ have already been reported in
\cite{B2pi_plb}. We note a small discrepancy that is accounted for by
the following: in
\cite{B2pi_plb} we used the fitted two-point function values of the
mass and overlap factor, whereas here we use the two-point function
itself to determine $Z_V^{\rm eff}$. This is expected to reduce the
statistical errors as the ratio is independent of $t_y$.

\begin{table}
\begin{center}
\caption{$Z_V^{\rm eff}$ and quark mass. $am_Q$ is the bare quark
	mass appearing in the current and $am_q$ is the bare quark
	mass of the light, spectator quark.}
\label{tab:ZV_VS_M}
\begin{tabular}{cccc}
\hline\hline
$\beta$ & $ am_Q$ & \multicolumn{2}{c}{$Z_V^{\rm eff}$}  \\\hline
        &             & $am_q=0.0332$ & $am_q=0.0195$ \\
        & $0.485$    & $1.3122^{+3}_{-3}$ & $1.3111^{+3}_{-3}$ \\
\raisebox{2.0ex}[0pt]{$6.2$}& $0.268$    & $1.0871^{+3}_{-3}$ & $1.0873^{+4}_{-4}$ \\\hline
        &             & $am_q=0.0502$ & $am_q=0.0298$ \\
        & $0.756$    & $1.5963^{+8}_{-8}$ & $1.5946^{+11}_{-13}$ \\
        & $0.566$    & $1.3876^{+6}_{-6}$ & $1.3843^{+\ 9}_{-10}$ \\
\raisebox{2.0ex}[0pt]{$6.0$}& $0.392$    & $1.2058^{+3}_{-4}$ & $1.2044^{+5}_{-5}$ \\
        & $0.231$    & $1.0333^{+3}_{-3}$ & $1.0333^{+3}_{-3}$ \\
\hline\hline
\end{tabular}
\end{center}
\end{table}

\subsection{The axial current}
The axial current is not conserved, and so we rely on HQS to determine
the effective renormalisation constant. The form factor $h_{A_1}$ at
zero recoil is equal to unity, up to symmetry breaking corrections,
which are reduced to $\order{1/m_Q^2}$ by Luke's theorem. This form
factor can be measured on the lattice, and is related to the effective
renormalisation by,

\begin{equation}
\label{eqn:ZA_eff}
Z_{A}^{\rm eff}h_{A_1}^{\rm latt}(1) = 1 + \beta_{A_1}(1)+{\mathcal O}(1/m_Q^2)\ .
\end{equation}
To measure the form factor, we do not rely on degenerate initial and
final states only. When they are not degenerate, the renormalisation
depends on the average mass of the quarks in the current. Our results
are shown in Table \ref{tab:ZA_VS_M}. Entries that have the repeated
average quark mass are taken from matrix elements with the initial and
final states interchanged.

\begin{table}
\begin{center}
\caption{$Z_A^{\rm eff}$ and average heavy quark mass.}
\label{tab:ZA_VS_M}
\begin{tabular}{cccc}
\hline\hline
$\beta$ & $\langle am_Q\rangle$ & \multicolumn{2}{c}{$Z_A^{\rm eff}$}  \\\hline
        &             & $am_q=0.0332$ & $am_q=0.0195$ \\
        & $0.485$    & $1.29^{+3}_{-3}$ & $1.27^{+4}_{-4}$ \\
        & $0.429$    & $1.26^{+3}_{-3}$ & $1.25^{+4}_{-3}$ \\
        & $0.377$    & $1.24^{+3}_{-3}$ & $1.22^{+4}_{-3}$ \\ 
	& $0.377$    & $1.23^{+3}_{-3}$ & $1.23^{+4}_{-3}$ \\ 
\raisebox{2.0ex}[0pt]{$6.2$}&$0.326$    & $1.20^{+3}_{-3}$ & $1.19^{+3}_{-3}$ \\ 
	& $0.321$    & $1.18^{+3}_{-3}$ & $1.17^{+4}_{-3}$ \\ 
	& $0.268$    & $1.10^{+3}_{-2}$ & $1.09^{+3}_{-3}$ \\ 
	& $0.218$    & $1.10^{+2}_{-3}$ & $1.09^{+3}_{-3}$ \\ \hline 
        &             & $am_q=0.0502$ & $am_q=0.0298$ \\
        & $0.756$    & $1.55^{+5}_{-5}$ & $1.52^{+7}_{-7}$ \\
        & $0.661$    & $1.45^{+5}_{-5}$ & $1.43^{+6}_{-6}$ \\
        & $0.661$    & $1.48^{+5}_{-5}$ & $1.46^{+7}_{-7}$ \\
        & $0.573$    & $1.40^{+5}_{-4}$ & $1.38^{+5}_{-6}$ \\
        & $0.573$    & $1.43^{+5}_{-5}$ & $1.42^{+7}_{-7}$ \\
        & $0.566$    & $1.37^{+4}_{-5}$ & $1.36^{+5}_{-6}$ \\
        & $0.493$    & $1.33^{+4}_{-4}$ & $1.31^{+5}_{-5}$ \\
	& $0.493$    & $1.41^{+5}_{-4}$ & $1.41^{+7}_{-7}$ \\
\raisebox{2.0ex}[0pt]{$6.0$}& $0.479$    & $1.32^{+4}_{-4}$ & $1.31^{+5}_{-5}$ \\
        & $0.479$    & $1.33^{+4}_{-4}$ & $1.32^{+6}_{-6}$ \\
        & $0.398$    & $1.27^{+4}_{-4}$ & $1.25^{+4}_{-5}$ \\	
        & $0.398$    & $1.20^{+4}_{-4}$ & $1.32^{+6}_{-6}$ \\
	& $0.391$    & $1.22^{+4}_{-4}$ & $1.21^{+5}_{-5}$ \\
	& $0.311$    & $1.19^{+4}_{-4}$ & $1.18^{+4}_{-4}$ \\	
	& $0.311$    & $1.23^{+4}_{-4}$ & $1.23^{+5}_{-5}$ \\
        & $0.231$    & $1.03^{+3}_{-3}$ & $1.01^{+4}_{-4}$ \\
\hline\hline
\end{tabular}
\end{center}
\end{table}

\subsection{Comparison with other determinations}
\begin{figure}
\begin{center}
\epsfig{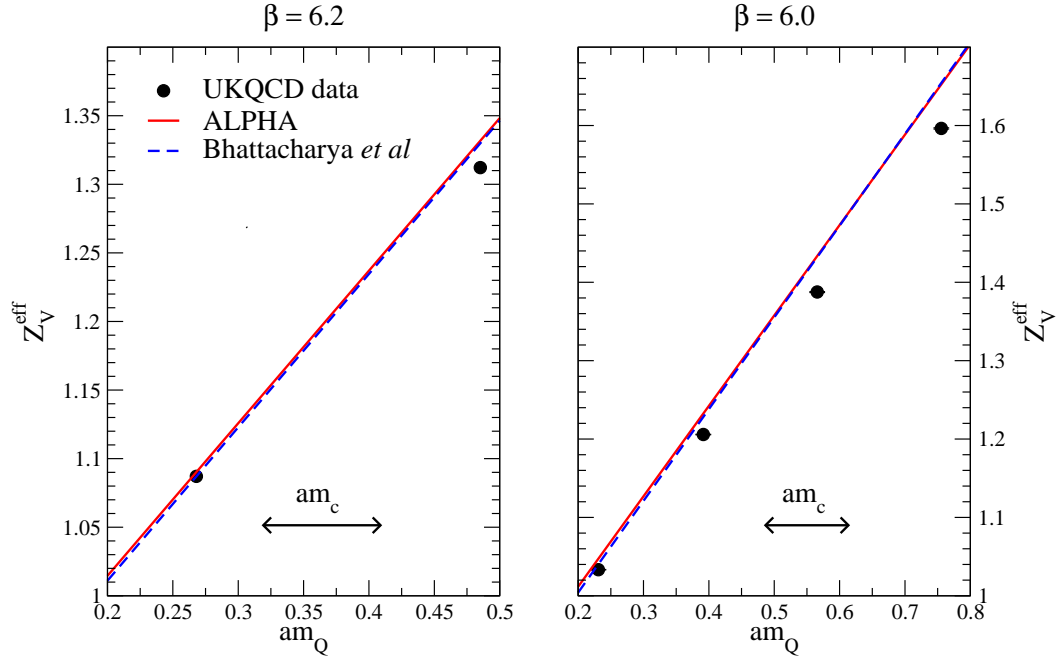}
\caption{$Z_V^{\rm eff}$ versus bare quark mass at both values of
$\beta$.  $am_c$ is the charm quark mass in lattice units, as
extracted from a lattice determination of the heavy-light
pseudoscalar meson mass. The range indicates the variation in the mass
due to the different choices of the scale-fixing quantity. }
\label{fig:ZV_VS_M}
\end{center}
\end{figure}

We compare our measurement of the effective renormalisation to
equation (\ref{eqn:eff_renorm}) using values of $Z_J$ and $b_J$
determined non-perturbatively by the ALPHA collaboration
\cite{alpha_np3_Z} and Bhattacharya {\em et
al}.~\cite{Bhatta_99,bhatta_plb,bhatta_2000}. This comparison for the
vector current is shown in Figure \ref{fig:ZV_VS_M} for the heaviest
spectator quark. The agreement is quite striking, especially
considering that the ALPHA collaboration determined the coefficients
at near zero quark mass and Bhattacharya {\em et al} for light quarks
(strange scale). Indeed, Bhattacharya {\em et al.} have already
compared their results at $\beta=6.2$ to those in~\cite{B2pi_plb}.

\begin{figure}
\begin{center}
\epsfig{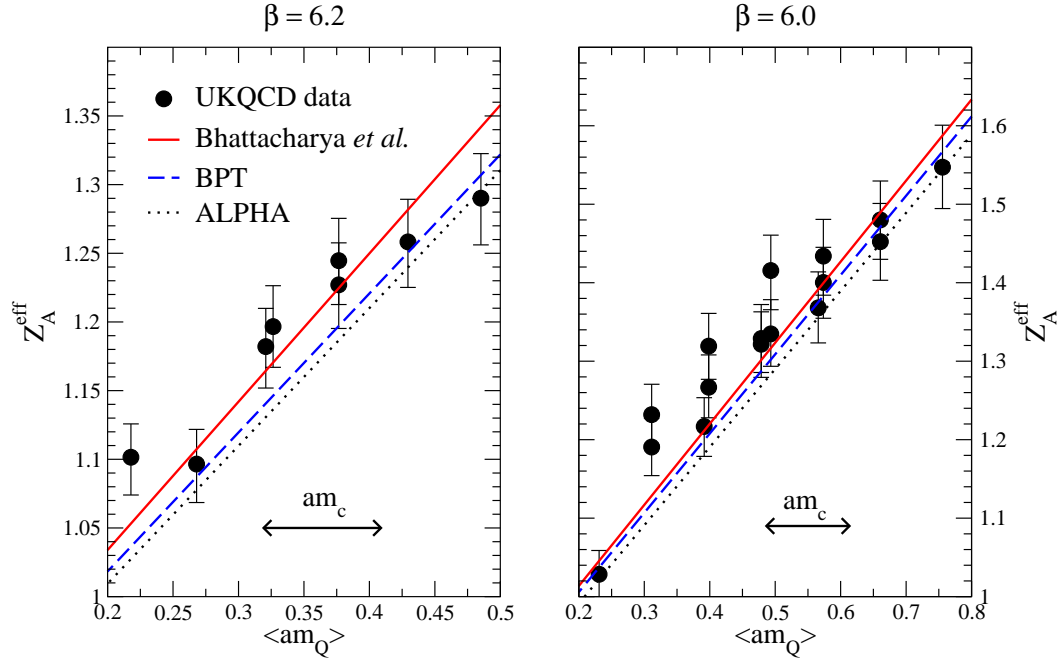}
\caption{$Z_{A}^{\rm eff}$ against average bare quark mass.}
\label{fig:ZA_VS_M}
\end{center}
\end{figure}

For the axial current, there is no NP determination of $b_A$ by the ALPHA 
collaboration. To facilitate a comparison, we use the one-loop value of 
$b_A$~\cite{alpha_np4_bA} 
\begin{equation}
  b_A=1+0.1522\ g^2
\end{equation}
with the ``boosted'' coupling, $g^2=g^2_0/u_0^4$, where the mean
link, $u_0$, is evaluated from the 4th root of the average
plaquette. This is shown in Figure \ref{fig:ZA_VS_M}. We also show the
values of $Z_A^{\rm eff}$ evaluated in boosted perturbation theory
(BPT). Again, our data compares very well to other determinations.

We don't use our results to determine $b_J$ and $Z_J$; however, the comparison
suggests that lattice artifacts are not spoiling the renormalisation.

\section{Extraction of the Isgur-Wise function}

The form factors that are protected from power corrections by Luke's theorem
are also used to extract the Isgur-Wise function:

\begin{eqnarray}
\label{eqn:wisgur}
\xi(\omega) & \simeq & \frac{h_+(\omega)}{1+\beta_+(\omega)}\ ,\\
\xi(\omega) & \simeq & \frac{h_{A_1}(\omega)}{1+\beta_{A_1}(\omega)}\ .
\end{eqnarray}

The power corrections for $h_+$ and $h_{A_1}$ have been estimated and
found to be consistent with zero in the range $1.0\le \omega \le 1.2$,
and therefore are neglected.  It is not possible to extract the
Isgur-Wise function from the form factors for which $\alpha_j=0$ as
they differ from zero only by power corrections. In principle it would
be possible to measure the power corrections in this way. However, the
poor quality of the data for the sub-leading form factors prevents
this.  Similarly, the $h_V$ form factor could not be used because its
power correction is found to be large, as expected ~\cite{neubert_wisgur_rad}.  

\begin{figure}
\begin{center}
\epsfig{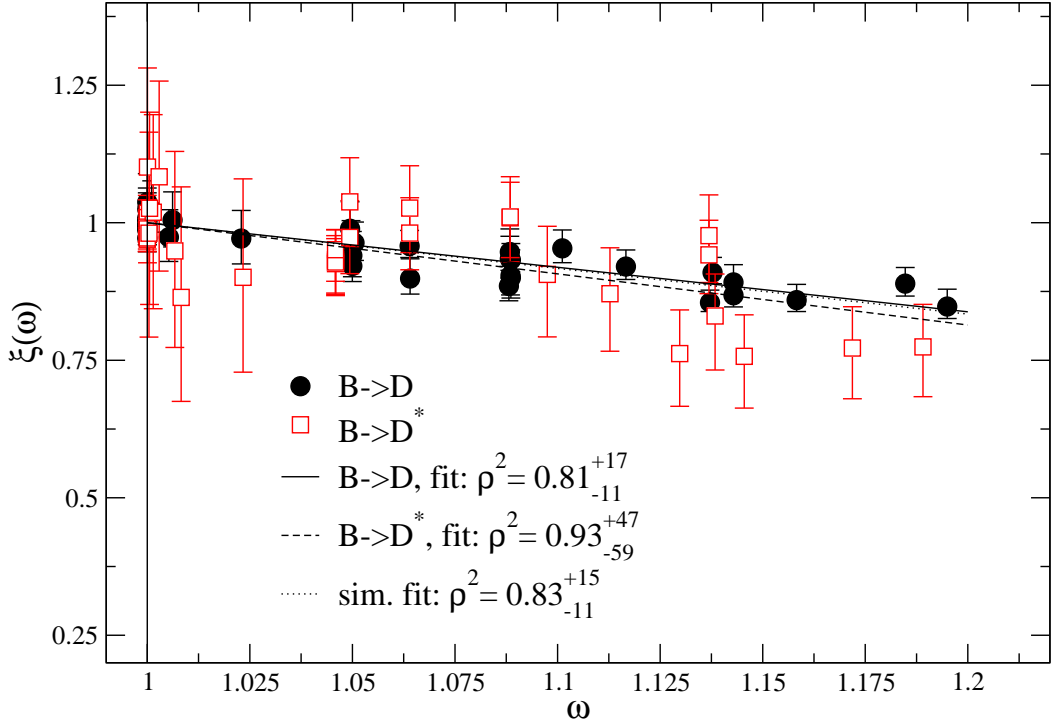}
\caption{The Isgur-Wise function from the $B\to D$ and $B\to D^*$ semi-leptonic decays,
with $\kappa_{\rm light}=0.1351$, including single-decay and simultaneous fits.}
\label{fig:wisgur}
\end{center}
\end{figure}

The Isgur-Wise function is fitted to the linear model
(\ref{eqn:linearwisgur}), at both values of the coupling. Simultaneous
fits of the Isgur-Wise function from the two distinct decays are also
performed; the assumption is made that the effect of mixing different
systematic errors (different decays) is no larger than the effect that
one gets by fitting simultaneously data from correlators with
different momenta. Results are summarised in Tables
\ref{tab:wisgur_fits_62},\ref{tab:wisgur_fits_60} and plotted in
Figure \ref{fig:wisgur}. The determination of the slope of the
Isgur-Wise function at $\beta=6.2$ from the simultaneous fit of the
two form factors, at the lightest value of the passive quark mass, is
taken as the central value.

\begin{table}
\begin{center}\caption{Results of the linear fits to the Isgur-Wise function,
        at $\beta=6.2$. Quoted errors are statistical.\smallskip}
\label{tab:wisgur_fits_62}
\begin{tabular}{@{}llcccccc}
\hline
&& \multicolumn{2}{c}{$B\to D$}
 &\multicolumn{2}{c}{$B\to D^*$}
 &\multicolumn{2}{c}{Sim. fit} 
\\
&$\kappa_{\rm light}$&$0.1346$ & $0.1351$ & $0.1346$ & $0.1351$ & $0.1346$ & $0.1351$ \\
\hline
&$\rho^2$ & $0.97^{+14}_{-9}$ &$0.81^{+17}_{-11}$ &$1.43^{+15}_{-32}$ &$0.93^{+47}_{-59}$
& $1.06^{+11}_{-11}$ & $0.83^{+15}_{-11}$\\
\hline
\end{tabular}
\end{center}
\end{table}

\begin{table}
\begin{center}\caption{Results of the linear fits to the Isgur-Wise function,
       at $\beta=6.0$. Quoted errors are statistical.\smallskip}
\label{tab:wisgur_fits_60}
\begin{tabular}{@{}llcccccc}
\hline
&& \multicolumn{2}{c}{$B\to D$}
 &\multicolumn{2}{c}{$B\to D^*$}
 &\multicolumn{2}{c}{Sim. fit} 
\\
&$\kappa_{\rm light}$&$0.13344$ & $0.13417$ & $0.13344$ & $0.13417$ & $0.13344$ & $0.13417$ \\
\hline
&$\rho^2$ & $0.94^{+25}_{-16}$ &$0.91^{+32}_{-22}$ &$0.68^{+15}_{-34}$ &$0.88^{+25}_{-40}$
& $0.88^{+19}_{-18}$ & $0.90^{+25}_{-25}$\\
\hline
\end{tabular}
\end{center}
\end{table}

\section{Quark mass dependence}

Close to the heavy quark limit ($m_b,m_c \to \infty$), at each fixed value
of $\omega$, the Isgur-Wise function does not depend on the mass of
the heavy quarks. However, the finite size of the simulated masses can
introduce a quark mass dependence, which has to be examined.

\subsection{Light quark mass dependence}

In the dataset used in this work, only two values of the passive quark
mass are available at each value of the coupling, making light quark
mass extrapolations impossible. The values of the form factors with
the two different light masses are compared and found to be
statistically consistent with each other.  However, the global fits of
the Isgur-Wise function with the two different passive quarks yield
slightly different, if still statistically compatible results. This
effect is taken into account as a systematic uncertainty on the
extracted slope of the Isgur-Wise function.

\begin{figure}
\begin{center}
\epsfig{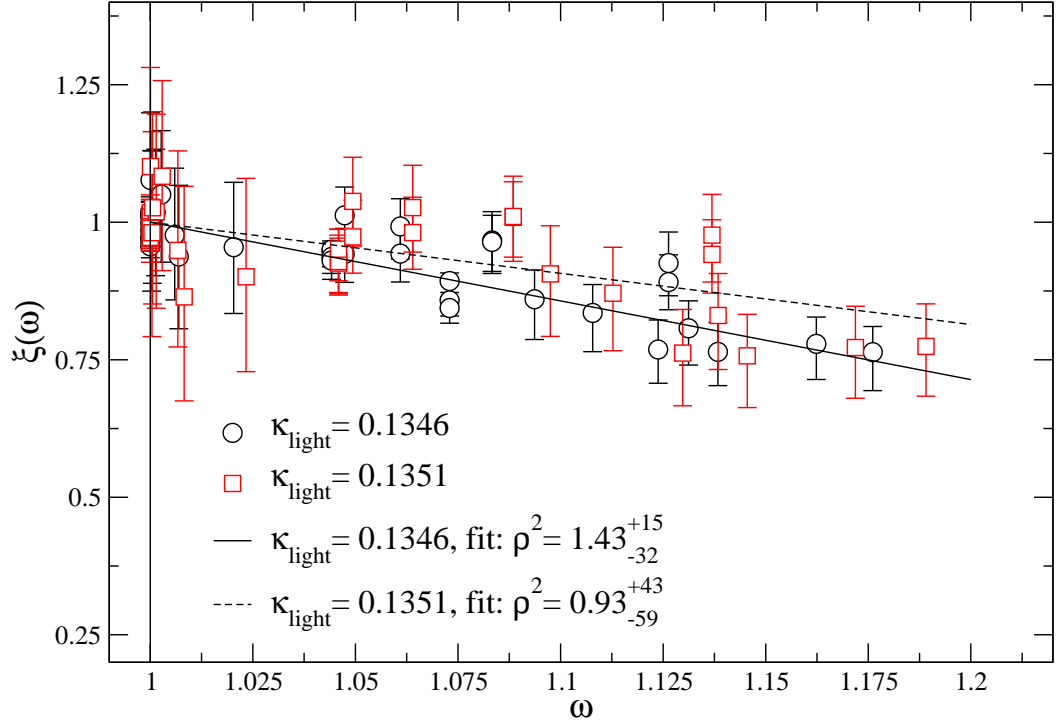}
\caption{The Isgur-Wise function from the $h_{A_1}$ form factor, for both values of
the light quark mass, at $\beta=6.2$.}
\label{fig:light_comp_hA1}
\end{center}
\end{figure}

\begin{figure}
\begin{center}
\epsfig{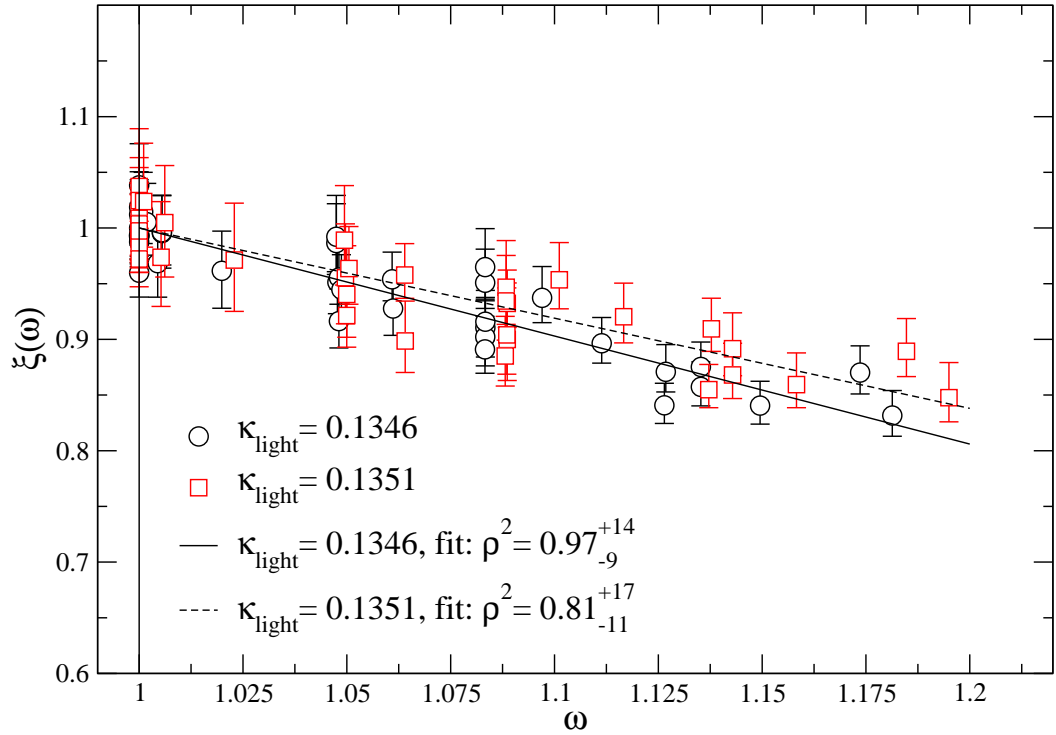}
\caption{The Isgur-Wise function from the $h_+$ form factor, for both values of
the light quark mass, at $\beta=6.2$.}
\label{fig:light_comp_hplus}
\end{center}
\end{figure}

Figures \ref{fig:light_comp_hA1} and \ref{fig:light_comp_hplus}
show a comparison of the fits of the Isgur-Wise function with the two
different values of the light quark mass, at $\beta=6.2$, for both
decays.

\subsection{\label{sec:HQM}Heavy quark mass dependence}

\begin{figure}
\begin{center}
\epsfig{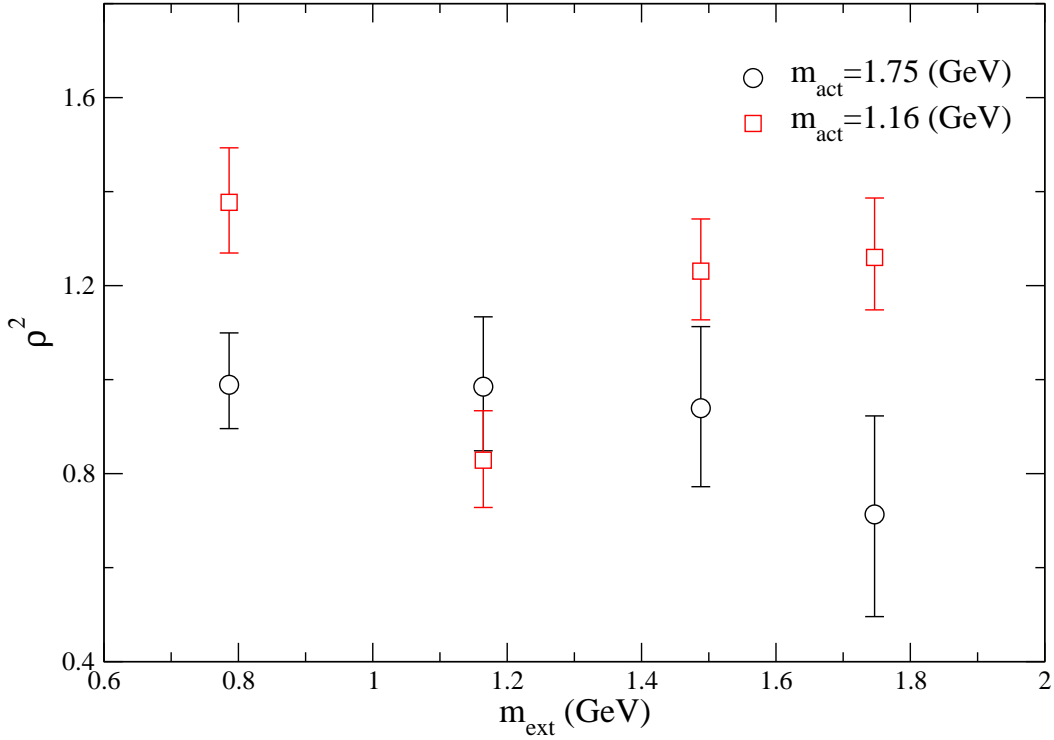}
\caption{Fits of the Isgur-Wise function at fixed values of the active quark propagator mass against the mass of the extended quark propagator, with $\kappa_{\rm light}=0.1346$, at $\beta=6.2$. Quoted quark masses are improved and renormalised according to (\ref{eqn:quark_mass}). }
\label{fig:wisgur_mass_3460}
\end{center}
\end{figure}

\begin{figure}
\begin{center}
\epsfig{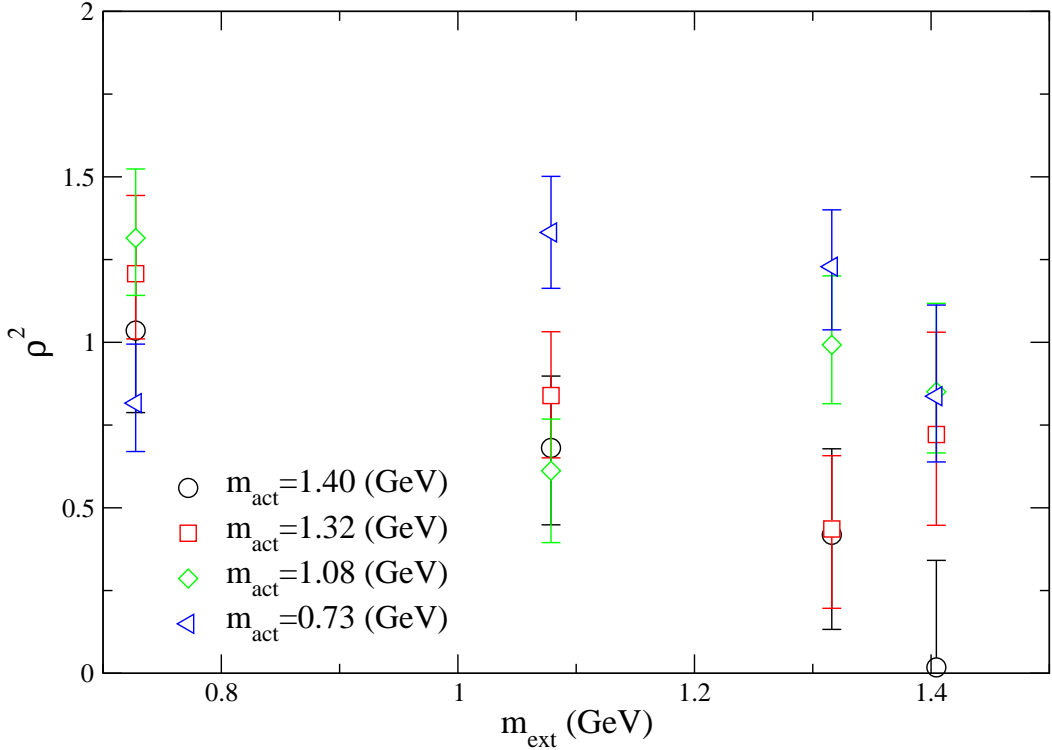}
\caption{Fits of the Isgur-Wise function at fixed values of the active quark propagator mass against the mass of the extended quark propagator, with $\kappa_{\rm light}=0.13344$, at $\beta=6.0$. Quoted quark masses are improved and renormalised according to (\ref{eqn:quark_mass}). }
\label{fig:wisgur_mass_3344}
\end{center}
\end{figure}

In this work the radiative corrections are applied, and the Isgur-Wise
function is evaluated in a range in $\omega$ in which the lattice
estimates of the power corrections for $h_+$ and $h_{A_1}$ are found
to be statistically consistent with zero.  The Isgur-Wise function is
extracted from the form factors by dividing off the radiative
corrections. This procedure should remove all residual quark mass
dependence at fixed $\omega$, if we are sufficiently close to the
heavy quark limit. To study any residual heavy quark mass dependence,
the Isgur-Wise function is extracted from all the momentum channels,
holding all the quark masses fixed. For each heavy quark mass
combination, the Isgur-Wise function is fitted, using all the
kinematic channels, and its slope is extracted. Then, for each value
of the extended quark mass, the result of the fit is plotted against
the value of the active quark mass. The results, for both values of
$\beta$, at fixed values of the light quark mass, are shown in Figures
\ref{fig:wisgur_mass_3460} and \ref{fig:wisgur_mass_3344}. As one
can see, there is no statistically significant heavy quark mass
dependence. It should be noted that fixing the quark masses
considerably reduces the number of points that are available to each
of the fits: consequently, some of the fits are quite poor.

\section{Ratios of form factors}

The two following ratios of form factors are also calculated:

\begin{eqnarray}
R_1(\omega)&=&\frac{h_V(\omega)}{h_{A_1}(\omega)}\ , \\
R_{2}(\omega)&=&\frac{h_{A_3}(\omega)+\frac{m_{D^*}}{m_B}h_{A_2}(\omega)}{h_{A_1}(\omega
)}\ .
\end{eqnarray}

\begin{figure}
\begin{center}
\epsfig{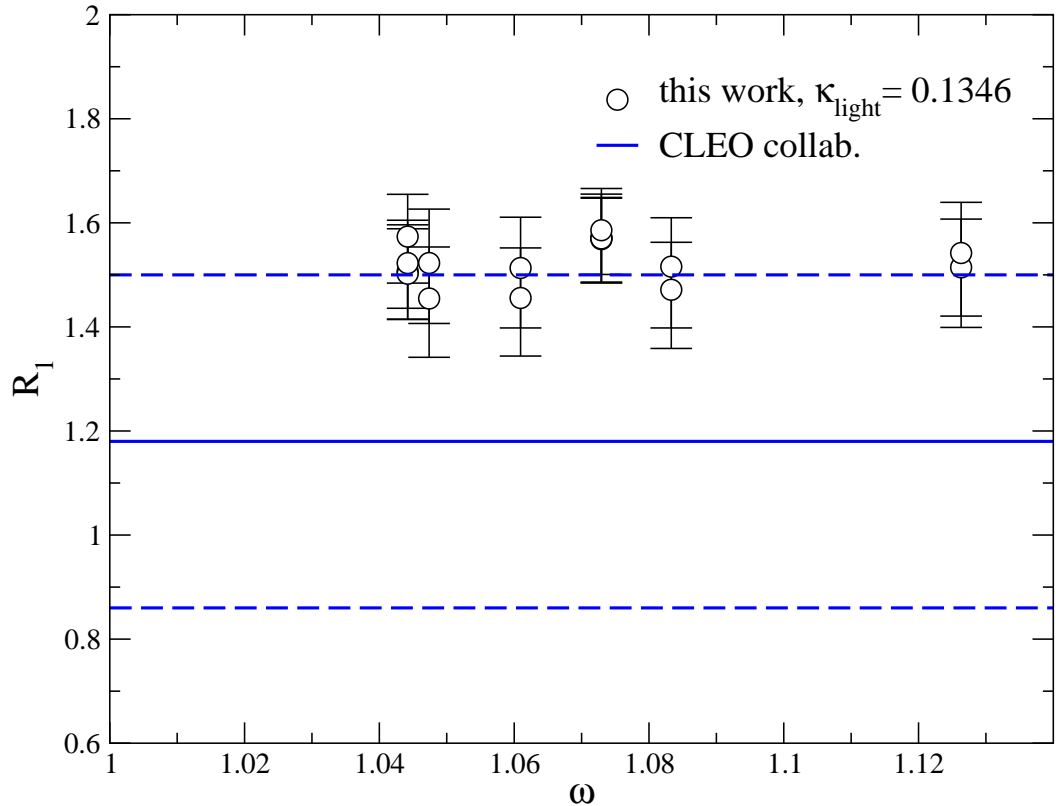}
\caption{The ratio $R_1$. The range of the experimental determination is shown by the dashed lines.}
\label{fig:R1}
\end{center}
\end{figure}

\begin{figure}
\begin{center}
\epsfig{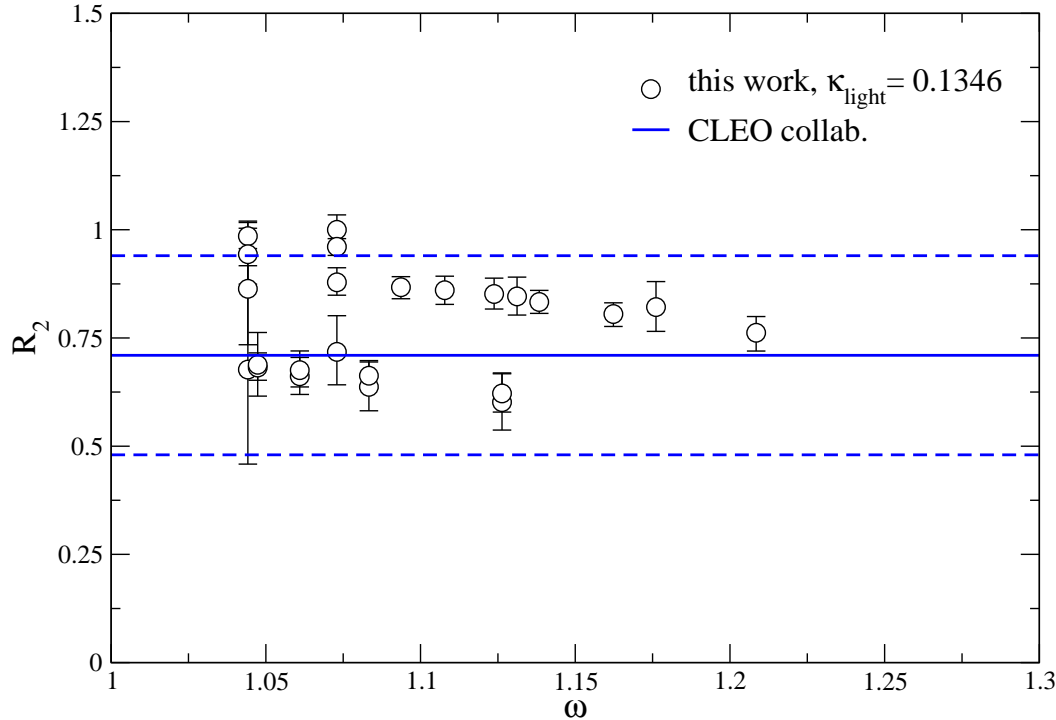}
\caption{The ratio $R_2$. The range of the experimental determination
is shown by the dashed lines. The data points are scattered in a wide range
because they correspond to different kinematic channels, with different systematic
effects.}
\label{fig:R2}
\end{center}
\end{figure}

These two ratios would be equal to one in the absence of symmetry
breaking corrections. Figures \ref{fig:R1} and \ref{fig:R2} show
that this work's determinations are in agreement with the experimental
determinations by the CLEO collaboration \cite{cleo_HLFF_ratios}, that
quote

\begin{eqnarray}
\label{eqn:CLEO_ratios}
R_1 & = & 1.18(32)\nonumber \\
R_2 & = & 0.71(23)\ .
\end{eqnarray}

In the case of $R_1$ there is some evidence of a systematic deviation
between our results and experiment. However, it should be noted that
this form factor has a strong dependence on the improvement
coefficient of the vector current, $c_V$, which is poorly known even
at $\beta=6.2$.

\section{Systematic uncertainties and comparison with other results}

\subsection{Systematic uncertainties}
Several sources of systematic uncertainty are
considered. Discretisation effects are estimated by comparing the
results at the two values of $\beta$. The results at $\beta=6.2$ are
taken as our central values. At $\beta=6.0$, the extraction of the
form factors is affected by considerably larger statistical
errors. This is reflected in the large statistical errors on the slope
of the Isgur-Wise function.

The systematic uncertainty induced by the scale fixing quantity is
evaluated by comparing fits of the Isgur-Wise function using $r_0$ and
$m_{\rho}$ to set the scale. In all cases, it is found to be an effect
smaller than $1\%$. This is mainly due to the fact that the radiative
corrections depend only logarithmically on the lattice spacing
(through quark masses). Furthermore, the effect of the definition of
the RGI quark mass and the values of the coefficients $Z_m$ and $b_m$
only enters through the radiative corrections.

The quark masses used in this calculation are around the charm quark
mass. It is necessary to check for any residual heavy quark mass
dependence beyond radiative and leading order power corrections. This
has been examined and no statistically significant trend has been
observed (see section \ref{sec:HQM}). No numerical estimate for this
is included in our systematic error.  The light quark masses are
around the strange quark mass, and thus the Isgur-Wise function
extracted is $\xi_s$ rather than $\xi$. We examine the light quark
mass dependence by comparing $\xi$ measured at the two values of quark
mass. An extrapolation is not attempted, but the lightest mass is
taken as a central value and the difference as a systematic
uncertainty.

The statistical errors on $h_{A_1}$ are much larger than those on
$h_+$ and so the latter form factor dominates the simultaneous
fit. Fitting to either form factor on its own produces the same result
within statistical error. In all cases, the data satisfy the HQS
condition $\xi(1)=1$ to such an extent that imposing this condition
on the fit has no sizeable effect.

It is rather difficult to quantify the effects of quenching other than
by performing the calculation in full QCD. There is some evidence that
quenching could have a sizeable effect in heavy-light decay
constants~\cite{Ryan_2001}, around $10-15\%$. However, the latter are
dimensionful quantities and the scale setting ambiguity in the
quenched approximation is sizeable. It is possible that large
unquenching effects in the decay constant are at least partly due to
the scale setting ambiguity~\cite{chrism_2001}. The Isgur-Wise
function is a dimensionless quantity and we have seen that it is
almost independent of the choice of the scale fixing quantity. There
is some evidence that quenching effects on the ratio $\xi_s/\xi_d$ are
mild~\cite{quench_HHCPT}.  It is reasonable to assume that quenching
effects will not be larger than any other systematic effect in this
calculation.

Our systematic uncertainties are then the difference of $\xi$ obtained
with the two values of the light quark mass ($+23\%$), with the two
values of the coupling ($+8\%$) and with the quantity used to set the
scale ($-1\%$). Positive and negative systematic uncertainties have
been separately added in quadrature.

Therefore, our final result is

\begin{equation}
\rho^2=0.83^{+15+24}_{-11-1}\ .
\end{equation}

\subsection{Comparison with other results}
Other recent lattice results for heavy-light to heavy-light decays
have concentrated on computing the value of the physical form factor
at zero recoil to extract $|V_{cb}|$. They use ratios of matrix
elements to cancel uncertainties~\cite{fnal_hplus} and further
calculate the HQS breaking corrections \cite{fnal_zero_recoil}.
Preliminary UKQCD results on this dataset for
$h_+$~\cite{douglas_wisgur} and $h_{A_1}$~\cite{gnl_wisgur} are
consistent with each other and this work.  Previous UKQCD
determinations~\cite{booth_wisgur,lellouch_wisgur,bowler_wisgur} give
$\rho^2=1.2^{+2}_{-2}{}^{+2}_{-1}$ with the spectator quark mass
around strange, and $\rho^2=0.9^{+2}_{-3}{}^{+4}_{-2}$ when
extrapolated to the chiral limit. Hashimoto and
Matsufuru~\cite{HM_wisgur} compute $\xi$ using lattice HQET, but with
an incomplete set of the $1/m_Q$ corrections, and quote
$\rho^2=0.70(17)$.
 
We also compare our result for the slope of the Isgur-Wise function
to experiment. Both CLEO and Belle have recent measurements of 
$B\to D^{\star}$ decays~\cite{CLEO_wisgur,Belle_wisgur} and quote
\ba
  \rho^2&=&1.67(11)(22) \ {\rm CLEO} \nonumber\\
  \rho^2&=&1.35(17)(19) \ {\rm Belle} 
\ea
Our result is significantly lower than the experimental ones, but
consistent with other quenched lattice determinations. However, the
experiments measure the slope of ${\mathcal F}(\omega)$, the physical
form factor, which is equal to the Isgur-Wise function only in the
infinite quark mass limit. Applying radiative corrections $\eta_A$ to
our results would increase the slope by around $5\%$. Considering the size
of our uncertainties, this effect is small and we neglect it for this
comparison.

\begin{ack}
The authors acknowledge PPARC grants PPA/G/O/$1998$/$00621$ \\ and 
PPA/G/0/$2000$/$00456$. RDK acknowledges grant
PPA/N/S/$2000$/$00217$, and CMM PPA/P/S/$1998$/$00255$. The authors thank
J.M.~Flynn and C.~McNeile for discussions.
This work was supported by the European Community's Human potential programme
under HPRN-CT-2000-00145 Hadrons/LatticeQCD
\end{ack}

\end{document}